%% file: scaling_obsarv.tex
\documentclass[twocolumn,superscriptaddress,showpacs,amssymb,amsmath,amsfonts,aps]{revtex4}

\date{\today }
\input{epsf}
\usepackage{epsfig}
\usepackage{latexsym}
\usepackage{amssymb}
\usepackage{graphicx}
\usepackage[dvips]{color}
\begin{document}
\title {
\huge {\bf Observation of Nuclear Scaling }   \\
\huge {\bf in the $A(e,e^{\prime} )$ Reaction at $x_B>$1 } 
} 

\input authorE2.tex
\begin{abstract}
The ratios of inclusive electron scattering cross sections of $^4$He,
$^{12}$C, and $^{56}$Fe to $^3$He have been measured for the first
time. It is shown that these ratios are independent of $x_B$ at
Q$^2>$1.4 (GeV/c)$^2$ for $x_B>$ 1.5 where the inclusive cross section
depends primarily on the high-momentum components of the nuclear wave
function. The observed scaling shows that the momentum distributions at 
high-momenta have the same shape for all nuclei and differ only by a scale factor.
The observed onset of the scaling at Q$^2>$1.4 and $x_B >$1.5 is 
consistent with the kinematical expectation that two nucleon short range 
correlations (SRC) are dominate the nuclear wave function at $p_m\gtrsim$ 
300 MeV/c. The values of these ratios in the scaling region can be
related to the relative probabilities of SRC in nuclei with A$\ge$3.  Our
data demonstrate that for nuclei with A$\geq$12 these probabilities
are 5-5.5 times larger than in deuterium, while for $^4$He it is
larger by a factor of about 3.5.
\end{abstract}
\maketitle
\vspace{1cm}

\normalsize{
\section {\large \bf INTRODUCTION} 
Due to the strong interaction and short distances between the nucleons
in nuclei, there is a 
significant probability for nucleon wave
functions to overlap, resulting in short range nucleon-nucleon
correlations (SRC) in nuclei \cite{FSreps}. Investigation of SRC is
important for at least  two reasons. First, because of the short range
nature of these correlations, they should contribute significantly to
the high-momentum component of the nuclear wave function. 
Second,  scattering from nucleons in
SRC will provide unique data on the modification of deeply bound
nucleons, which is extremely important for a complete understanding of
nucleon structure in general.

High-energy inclusive electron scattering from nuclei, $A(e,e')$, is
one of the simplest ways to investigate SRC. In particular, it is
probably the best way to measure the  probabilities of SRC in
nuclei.  The main problem in these studies is selecting the electron-SRC
scattering events from the  orders-of-magnitude larger
background of inelastic and/or quasi-elastic interaction of electrons
with the uncorrelated low-momentum nucleons.
 
By measuring cross sections at 
\begin{equation}
x_B=\frac {Q^2}{2M\nu }>1,
\label{xm1}
\end{equation}
contributions from inelastic electron-nucleon scattering and meson
exchange currents (at high Q$^2$) can be significantly reduced, which corresponds 
to studying the low-energy-loss side of the quasi-elastic peak.
In Eq.~\ref{xm1} Q$^2$ is the four-momentum squared of the virtual
photon (Q$^2 = -q^\mu q_\mu >$0), $\nu$ is the energy transfer, 
$x_B$ is the Bjorken scaling variable, and $M$ is the nucleon mass,

Many previous analyses of data in this kinematic region concentrate on
using $y$-scaling to study nucleon momentum distributions (see {\it e.g.}
Refs.~\cite{TWD99}, ~\cite{Arrington99}). While this technique provides some
information about the nuclear wave function, it does not measure the
probability of finding SRC in nuclei.

Meanwhile, the data at $x_B >$1 can be used to measure the probability of
finding SRC in nuclei.  There are theoretical predictions that 
at momenta higher than the Fermi momentum, 
the nucleon momentum distributions
in light and heavy nuclei are similar (see {\it e.g.} Ref.~\cite{Ciofi}). This
implies that they originate predominantly from the interaction between
two nearby nucleons, {\it i.e.} due to SRC. 
If the $A(e,e')$ cross section depends primarily on the nuclear wave function and
the shape of this wave function at high-momentum is really  universal,
then in this high-momentum region the ratio of weighted $(e,e')$ cross 
sections for different nuclei
 \footnote {Hereafter, by the
ratio of the cross sections we will mean the ratios of the cross sections 
weighted by $A$. We will separately discuss effects due to $\sigma_{ep} >
\sigma_{en}$ that are important for $^3$He due to $Z$ not equal to $N$.}
should scale, {\it i.e.} they 
should be independent of electron scattering variables
(Q$^2$ and $x_B$), with the magnitude of the scaling factor being
proportional to the relative probability of SRC in the two nuclei
\cite{FSworld,FSDS}.

In Ref.~\cite{FSDS} this was checked by analyzing existing SLAC
$A(e,e')$ data for deuterium \cite{Schutz,Rock,Arnold} and heavier
nuclei \cite{Day1}.  They found an indication of scaling at Q$^2>$1
and $x_B \ge $1.5. However, since the data for deuterium and the heavy
nuclei were collected in different experiments at similar Q$^2$ but at
different electron scattering angles and incident electron energies,
to find the ratios at the same values of ($x_B,Q^2$), a complicated
fitting and interpolation procedure was applied \cite{FSDS} to the
data.  The main problem was that the cross sections varied very
strongly with angle, incident energy, and Q$^2$.  To simplify the
interpolation, the electron-deuteron cross section was first divided
by the theoretical calculation within the impulse approximation.
Therefore, the data are not purely experimental, since they include
the theoretical calculations, and the ratios may have been affected by
the fitting and interpolation procedures.

In this work, the yields of the reaction $A(e,e')$ for $^3$He, $^4$He, $^{12}$C, and
$^{56}$Fe targets are measured in the same kinematical conditions, and the
ratios $A(e,e')$/$^3$He$(e,e')$ are obtained for 1$< x_B<$ 2 and Q$^2>$0.65
(GeV/c)$^2$. Furthermore, using the scaling behavior of these ratios,
the relative probability of $NN$ SRC for the various nuclei have been extracted.

\section {\large \bf Kinematics and Predictions} 

In order to suppress the background from quasi-elastic interactions of
electrons with the uncorrelated low-momentum nucleons (see
Fig.~\ref{fig:diagr}a), we further restrict the kinematic variables $x_B$
and Q$^2$.
\begin{figure}[ht]
\vspace{35mm} 
\centering{\includegraphics{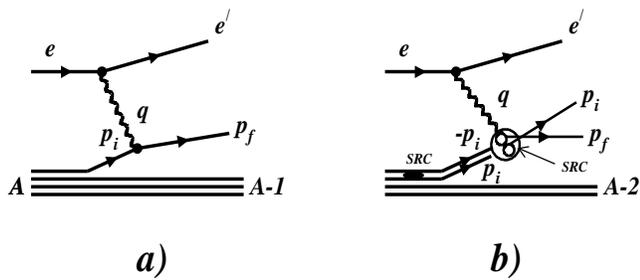}}
\caption[]{Two mechanisms of $A(e,e')$ scattering. a) single nucleon model; 
b) Short Range Correlation model.}
\label{fig:diagr}
\end{figure}

For quasi-elastic $A(e,e')$ scattering, $x_B$, Q$^2$, and the minimum
$A-1$ recoil momentum contributing to the reaction are related by
energy and momentum conservation:
\begin{eqnarray}
(q+p_A-p_{A-1})^2 = p^2_f = m^2_N, 
\label{QE1}
\end{eqnarray}
where $q$, $p_A$, $p_{A-1}$, and $p_f$ are the four-momenta of the virtual photon,
target nucleus, residual $A-1$ system, and knocked-out nucleon
respectively (note that only $q$ and $p_A$ are known.)  From  Eq.(\ref{QE1}) 
one obtains:
\begin{eqnarray}
\Delta M^2 - Q^2 + \frac{Q^2}{m_N
x_B}\left(M_A-\sqrt{M_{A-1}^2+ \vec p_m{}^2}\right) 
\nonumber \\ 
- 2\vec q \cdot\vec p_m - 2M_A\sqrt{M_{A-1}^2+\vec p_m{}^2}~=~0,
\label{QE2}
\end{eqnarray}
where $\Delta M^2 = M_A^2 + M_{A-1}^2 - m_N^2$ and $\vec p_m = \vec
p_f - \vec q = -\vec p_{A-1}$ is the recoil momentum involved in the
reaction (sometimes referred to as the `missing momentum' in $(e,e'p)$
reactions). Eq. (\ref{QE2}) defines a simple relationship between $\vert\vec
p_m{}^{min}\vert$ and $x_B$ at fixed Q$^2$. This relation for deuterium at
various values of Q$^2$ is shown in Fig. \ref{fig:PMissMin}a. Fig.
\ref{fig:PMissMin}b shows the same relationship for various nuclei at 
Q$^2 =2$ (GeV/c)$^2$. Note that
this relationship is different for the different nuclei, due primarily
to differences in the mass of the recoil $A-1$ system. 
This minimum recoil momentum is one of
the possible definitions of the scaling variable $y$.

\begin{figure}[ht]
\begin{center}
\epsfig{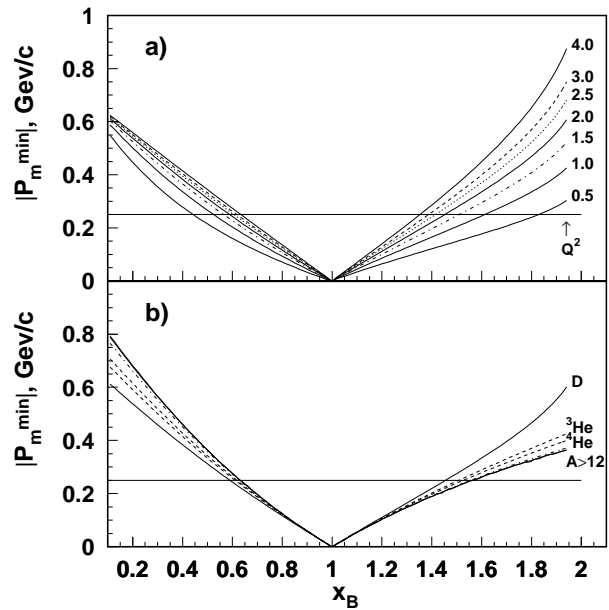}
\caption[]{The minimum  recoil momentum as a function of
$x_B$. a) For deuterium at several Q$^2$ (in (GeV/c)$^2$); b) For
different nuclei at Q$^2$ = 2.0 (GeV/c)$^2$. Horizontal lines at 250
MeV/c indicate the Fermi momentum typical of the uncorrelated
motion of nucleons in nuclei.}
\label{fig:PMissMin}
\end{center}
\end{figure}
One can see from Fig.\ref{fig:PMissMin} that for any nucleus $A$ and
fixed Q$^2$, we can find the value $x^o_B$ such that at $x_B~ > ~x^o_B$
the magnitude of the minimum recoil momentum, $\vert\vec
p_m{}^{min}\vert$, contributing to the reaction, exceeds
the average Fermi momentum in nucleus $A$.

It should be pointed out that the initial momentum of the struck nucleon 
$\vec p_i$ is  equal to $\vec p_m$ only in the simplest model where the
virtual photon is absorbed on one nucleon and that nucleon leaves the
nucleus without further interactions (the Plane Wave Impulse
Approximation). In reality, the $(e,e')$ reaction effectively
integrates over many values of $p_m \ge p_m^{min}$.  In addition, this
simple relation between recoil momentum and initial momentum is
modified by Final State Interactions (FSI) and the excitation energy
of the residual nucleus.  These make it difficult to determine the
nuclear wave function directly from $(e,e')$ cross sections.  However,
for our purposes, it is sufficient to know that when the minimum
recoil momentum contributing to the reaction is much larger than the
Fermi momentum, the initial momentum of the struck nucleon will also
be larger.

Let us now consider various predictions of the  ratios of weighted $(e,e')$
cross sections for different nuclei.  In the mechanism for inclusive
$(e,e')$ scattering at $x_B>$1 with virtual photon absorption on a
single nucleon and the $A-1$ system recoiling intact without FSI (see
Fig.~\ref{fig:diagr}a), the minimum recoil momentum for
different nuclei at fixed Q$^2$ differs and this difference increases
with $x_B$ (see Fig.~\ref{fig:PMissMin}).  Therefore, the cross
section ratio between different nuclei will increase with $x_B$ and
will not scale.

In the Short Range Correlations model of Frankfurt and Strikman
\cite{FSreps} (see  Fig.~\ref{fig:diagr}b) the high-momentum part of 
the nuclear momentum distribution is due to
correlated nucleon pairs.  This means that when the electron scatters
from a high-momentum nucleon in the nucleus, we can consider this
scattering as an electron-deuterium interaction with the spectator $A-2$
system almost at rest.  Therefore, according to 
Fig. \ref{fig:PMissMin}a, starting from some threshold 
$x_B^0$ for fixed Q$^2$ the cross section ratio
\begin{equation}
R^{A1}_{A2}(Q^2,x_B) = \frac {\sigma_{A_1}(Q^2,x_B)/A_1}{\sigma _{A_2}(Q^2,x_B)/A_2}, 
\label{ratio1a}
\end{equation}
where $\sigma _{A_1}$ and $\sigma _{A_2}$ are the inclusive electron
scattering cross sections from nuclei with atomic numbers $A_1$ and
$A_2$ respectively, will scale (will be constant). 
Scaling results from the dominance of SRC in the
high-momentum component of the nuclear wave function, and it should be
observed, for example, for the cross section ratios of heavy nuclei to light 
nuclei such as $^3$He.

Fig. \ref{fig:SRCPredict}a shows $R^C_{^3He}$, for $A_1=~^{12}$C and
$A_2=~^3$He, as a function of $x_B$ for Q$^2$ from 1.5 to 2.5
(GeV/c)$^2$ calculated in the SRC model \cite{Misak2}. The ratio for
$A_1=~^{56}$Fe and $A_2=~^3$He is shown in Fig. \ref{fig:SRCPredict}b. 
The calculations used the Faddeev wave function for $^3$He calculated using the
Bonn $NN$ potential~\cite{Bonn}. The momentum
distributions for heavier nuclei have been modeled through a
two component of momentum distribution using mean field distributions for
small nucleon momenta and using the deuteron momentum distribution
for $p > $250 MeV/c, scaled by factor $a_2(A)$,
per-nucleon probability of $NN$ SRC in nucleus A, estimated from
Ref.~\cite{FSDS}. The mean field momentum
distributions used the Harmonic Oscillator wave function for $^{12}$C
and the quasi-particle Lagrange Method of \cite{ZS} for $^{56}$Fe.
For the description of the $eN$
interaction, the 
inelastic form factor parameterization of Ref.~\cite{Bodek} and the dipole 
elastic form-factors have been used. These calculations are in reasonable 
agreement with existing $A(e,e')X$ experimental data from SLAC~\cite{Day2} 
and from Jefferson Lab Hall C~\cite{Petit}.

The ratios in Fig. \ref{fig:SRCPredict} show a nice plateau starting from
$x_B>$1.5 for both nuclei and all Q$^2$. The experimentally obtained ratio in the
scaling region can be used to determine
the relative probability of finding correlated $NN$ pairs in different
nuclei. However one needs to take into account two
main factors: first the final state interactions of a nucleon with the
residual system, and second the $NN$ pair center-of-mass motion.

In the SRC model, FSI do not destroy the
scaling behavior of the ratio, $R$. Indeed, in
the light-cone approximation of the SRC model, if the invariant mass
of the final $NN$ system is sufficiently large, 
$\sqrt{(q+m_D)^2}-m_D>$ 50-100 MeV, then the scattering amplitude will 
depend mainly on the light-cone fraction of the interacting nucleon's 
momentum $\alpha = (E- p_z)/M$, and has only a weak dependence on the 
conjugated variables $E + p_z$ and $p_t$ \cite{FSDS,FSS97,Misak1}. 
As a result, the closure approximation can be applied in the light-cone
reference frame, allowing us to sum over all final  states and use 
the fact that this sum is normalized to unity. After using the closure 
approximation the  inclusive cross section will depend on the light-cone
momentum distribution of the nucleon in the nucleus, integrated over the
transverse momentum of the nucleon, $\rho_A(\alpha)$ \cite{FSworld}. Thus, 
within the light cone description the Eq.(4) measures the ratio of 
$\rho_A(\alpha)$ for nuclei $A_1$ and $A_2$ in the high-momentum range 
of the target nucleon.

In the lab frame description (in the virtual nucleon approach), however, the
closure approximation cannot be applied for large values of interacting
nucleon momenta, and FSI should be calculated explicitly (see
{\it e.g.} Ref.~\cite{FSS97}). Within the SRC model at high recoil
momenta, FSI are dominated by the rescattering of the knocked-out
nucleon with the correlated nucleon in the SRC
\cite{FSDS,FSS97}. Therefore, FSI will be localized in SRC, and 
will cancel in the ratio $R$. 
As a result, Eq.(4) at $x_B>x^0_B$ could be related to 
the ratio of high-momrntum part of nucleon-momentum distributions in 
$A_1$ and $A_2$ 
nuclei ~\cite{FSS97}.

Having an underlying model of the nuclear spectral functions, one can
relate the
measured ratios in Eq.~(4) to the SRC properties of the nuclear wave function.
Within the spectral function  model \cite{FSreps}, in which correlated nucleon pair 
is assumed at rest with the nucleon momentum distribution in pair identical to that 
in deuteron, the ratio in Eq.~(4) could be directly related to 
the per-nucleon SRC probability in nucleus $A$ relative to
deuterium, $a_2(A)$. In models of the nuclear 
spectral function~\cite{CSFS} in which two-nucleon correlations are moving 
in the mean field of the spectator $A-2$ system, the analysis of Eq. (4) will 
yield slightly smaller values for $a_2(A)$.  Calculations by Ciofi degli
Atti~\cite {ciofipc} indicate that this motion does not affect the scaling but
can decrease the extracted $a_2(A)$ for $^{56}$Fe by up to 20\%.
However it is important to 
emphasize that since both approximations predict 
similar (light cone) momentum distributions, both models lead to a
similar ratio 
of the light-cone spectral functions and  the overall probability of high-momentum 
nucleons remains practically the same.


One can summarize the predictions of the SRC model for the ratios of
the inclusive cross sections from different nuclei as follows (see
Fig. \ref{fig:SRCPredict}):
\begin{itemize}
\item  Scaling ($x_B$ independence) is expected for Q$^2\ge$1
         (GeV/c)$^2$ and  $x_B^0\le x_B <$2 where $x_B^0$ is the
        threshold for high recoil momentum.
\item No scaling is expected for  Q$^2 <$ 1 (GeV/c)$^2$. 
\item For $x_B\le x_B^0$ the ratios should have a minimum at $x_B$ = 1
and should grow with $x_B$ since heavy nuclei have a broader momentum
distribution than light nuclei for $p < $0.3 GeV/c.
\item The onset of  scaling  depends on Q$^2$; $x_B^0$ should
        decrease with increasing Q$^2$.
\item In the scaling regime,  the ratios should be independent
      of Q$^2$. 
\item In the scaling regime the ratios should depend  only  weakly on $A$
         for  $A\ge 10$.  This reflects nuclear saturation.
\item Ratios in the scaling regime are proportional to the ratios of the two-nucleon SRC
probabilities in the two nuclei.
\end{itemize}

\begin{figure}[ht]
\begin{center}
\epsfig{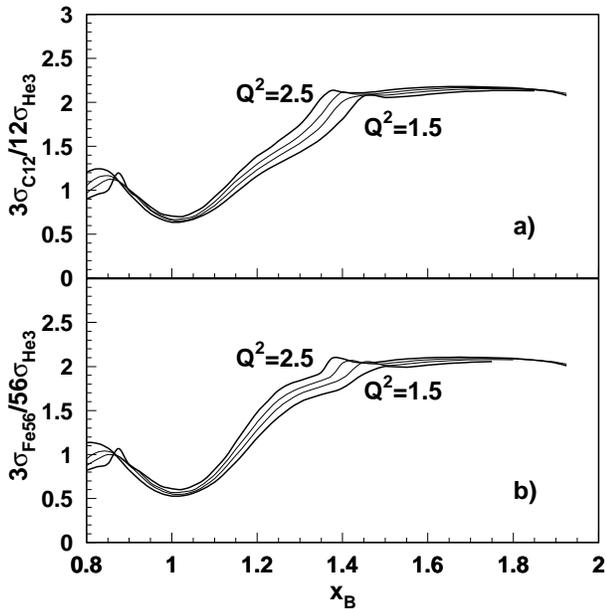}
\caption[]{SRC Model predictions for the normalized inclusive cross
section ratio as a function of $x_B$ for several values of Q$^2$ (in
(GeV/c)$^2$). Note the scaling behavior predicted for $x_B >$1.4.  a)
$^{12}$C to $^3$He, b) $^{56}$Fe to $^3$He.}
\label{fig:SRCPredict}
\end{center}
\end{figure}
Another possible mechanism for inclusive $(e,e')$ scattering at $x_B>$1
is virtual photon absorption on a single nucleon followed by $NN$
rescattering \cite{Benhar1,Benhar2}.  Benhar {\it et al.} \cite{Benhar1} 
use the nuclear spectral function in the lab system and calculate the FSI
using a correlated Glauber approximation (CGA), in which the initial
momenta of the re-scattered nucleons are neglected. In this model the cross
section at $x_B>$1 originates mainly from FSI and therefore the cross
section ratios will not scale.  This model predicts that these ratios
also depend on Q$^2$, since it includes a noticeable reduction of FSI
in order to agree with the data at Q$^2\ge$2 (GeV/c)$^2$.  Benhar
{\it et al.}  attribute this reduction in FSI to color transparency
effects \footnote{So far no color transparency effects are observed in
$A(e,e'p)X$ reactions at Q$^2\le$ 8 (GeV/c)$^2$~\cite{Garrow}.}.  The
requirement of large color transparency effects also results in a
strong $A$ dependence of the ratio since the amount of the FSI
suppression depends on the number of nucleons participating in the
rescattering.

The main predictions  of  the CGA model for the 
nuclear cross section ratios are as follows:
\begin{itemize}
\item No scaling is predicted for  Q$^2\ge$1 (GeV/c)$^2$ and
      $x_B <$2.
\item The nuclear ratios should vary with Q$^2$.
\item The ratios should depend on  $A$. 
\item The model is not applicable below Q$^2$=1 (GeV/c)$^2$.
\end{itemize}

Thus, measuring the ratios of inclusive $(e,e')$ scattering at $x_B>$1
and Q$^2>$1 (GeV/c)$^2$ will yield important information about the
reaction dynamics.  If scaling is observed, then the dominance of the SRC in
the nuclear wave function is manifested and the measured
ratios will contain information about the probability of two-nucleon
short range correlations in nuclei.

\section {\large \bf EXPERIMENT}
In this paper we present the first experimental studies of ratios of
normalized, and acceptance- and radiative-corrected, inclusive yields of
electrons scattered from $^4$He, $^{12}$C, $^{56}$Fe, and $^3$He measured
under identical kinematical 
conditions. 

The measurements were performed with the CEBAF Large Acceptance
Spectrometer (CLAS) in Hall B at the Thomas Jefferson National
Accelerator Facility. This is the first CLAS experiment with nuclear
targets. Electrons with 2.261 and 4.461 GeV energies incident on $^3$He,
$^4$He, $^{12}$C, and $^{56}$Fe targets have been used. 
We used helium liquefied in
cylindrical target cells 1 cm in diameter and 4 cm long, positioned
on the beam approximately in the center of CLAS.  
The solid targets were thin foils of $^{12}$C (1mm) and $^{56}$Fe (0.15 mm)
positioned 1.5 cm downstream of the exit window of the liquid
target. Data on solid targets have been taken with an empty
liquid target cell.

The CLAS detector \cite{CDR} consists of six sectors, each
functioning as an independent magnetic spectrometer. Six
superconducting coils generate a toroidal magnetic field primarily in
the azimuthal direction.  Each sector is instrumented with multi-wire
drift chambers \cite{Mestayer} and time-of-flight scintillator
counters \cite{Smith} that cover the angular range from 8$^{\circ}$ 
to 143$^{\circ}$, and,
in the forward region (8$^{\circ}<\theta<$45$^{\circ}$), with gas-filled threshold
Cherenkov counters (CC) \cite{Adams} and lead-scintillator sandwich-type
electromagnetic calorimeters (EC) \cite{Amarian}.  Azimuthal coverage
for CLAS is limited only by the magnetic coils, and is approximately
90\% at large polar angles and 50\% at forward angles.
The CLAS was triggered on
scattered electrons by a CC-EC coincidence at 2.2 GeV and by the
EC alone with a $\approx 1$ GeV electron threshold at 4.4 GeV.

For our analysis, electrons are selected in the kinematical region Q$^2
> $0.65 (GeV/c)$^2$ and $x_B>$1 where the contribution from the 
high-momentum components of the nuclear wave function should be enhanced.  
\begin{figure}[ht]
\begin{center}
\epsfig{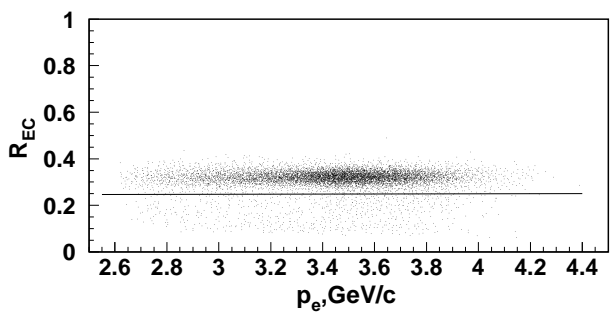}
\caption[]{The ratio (R$_{EC}$) of energy deposited in the CLAS
electromagnetic calorimeter (EC) to the
electron momentum $p_e$ as a function of $p_e$ at beam energy 4.461 GeV. 
The line at R$_{EC}\approx$ 0.25 is located 3 standard deviations below
the mean, as determined by measurements at several values of $p_e$.
This cut was used to identify electrons.}
\label{fig:REC}
\end{center}
\end{figure}
We also require that the energy transfer, $\nu$, should be $>$300 MeV 
(the characteristic missing energy for SRC is $\sim260$ MeV \cite{FSreps}). 
In this region one expects 
that inclusive  $A(e,e')$ scattering will proceed through the interaction of 
the incoming electron  with  a correlated nucleon in a SRC.

\subsection {\large \bf Electron Identification}

Electrons were selected in the fiducial region of the CLAS sectors.
The fiducial region is a region of azimuthal angle, for a given momentum
and polar angle, where the electron detection efficiency is constant.
Then a cut on the ratio of the energy deposited in the EC to the
measured electron momentum $p_e$ (R$_{EC}$) was used for final selection.
In Fig. \ref{fig:REC} R$_{EC}$
vs. $p_e$ for the $^{56}$Fe target at 4.4 GeV is shown.  
The lines shows the applied cut at $R_{EC}\approx 0.25$ which is located 3
standard deviations below the mean as determined by measurements at
several values of $p_e$.  A Monte Carlo simulation showed that these cuts
reduce the $A(e,e^{\prime})$ yield by less then 0.5\%.

\begin{figure}[ht]
\begin{center}
\epsfig{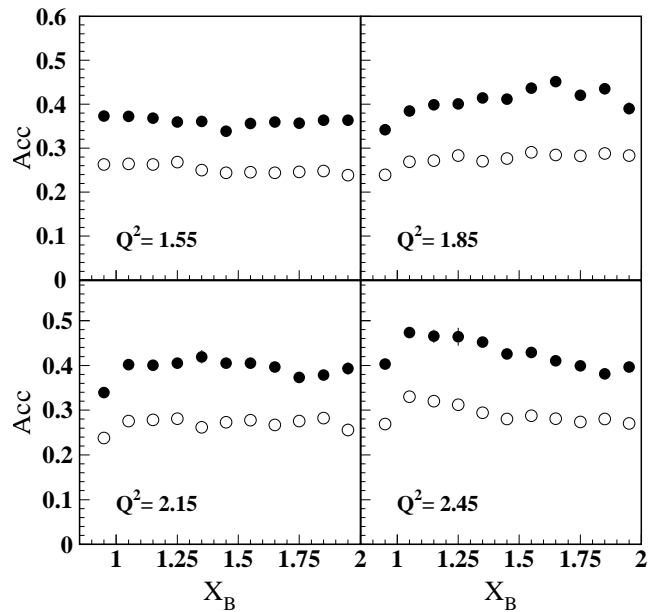}
\caption[]{The acceptance correction factors as a function of ${\it
x_B}$. $\bullet$  $^3$He,  $\circ$ $^{12}$C. Q$^2$ are in 
(GeV/c)$^2$.} 
\label{fig:accept}
\end{center}
\end{figure}

We estimated the $\pi^-$ contamination in the electron sample for a
wide angular range using
the photo-electron distributions in the CLAS Cherenkov counters. 
We found that for $x_B>$1 this is negligible.

\subsection{\large \bf Acceptance Corrections}
We used the Monte Carlo techniques to determine the electron acceptance
correction factors. Two iterations were done to optimize the cross
section model for this purpose.  In the first iteration we generated
events using the SRC model
\cite{Misak2} and determined the CLAS detector response using 
the GEANT-based CLAS simulation program, taking into account 
``bad'' or ``dead'' hardware
channels in various components of CLAS, as well as realistic position
resolution for the CLAS drift chambers. We then used the CLAS data analysis
package to reconstruct these events using the same electron
identification criteria that was applied to the real data. The
acceptance correction factors were found as the ratios of the number
of reconstructed  and simulated events in each kinematic bin. 
Then the acceptance corrections were applied to
the data  event-by-event, i.e. each event was weighted by the
acceptance factor obtained for the corresponding $(\Delta x_B,\Delta$Q$^2)$ 
kinematic bin and the cross sections were calculated as a function 
of $x_B$ and Q$^2$. For the second iteration the obtained cross sections were
fitted and the fit-functions 
were used to generate a new set of data, and the process was repeated.
Fig.~\ref{fig:accept} shows the electron acceptance factors after the second 
iteration for liquid ($^3$He) and solid ($^{12}$C) targets. We used the 
difference between the iterations as the uncertainty in the acceptance 
correction factor. Note that the acceptance for the carbon target is
smaller than for the helium target. This is due to the closer location of the
solid targets to the CLAS coils, which limit azimuthal angular coverage
of the detectors.
%
\begin{figure}[ht]
\begin{center}
\epsfig{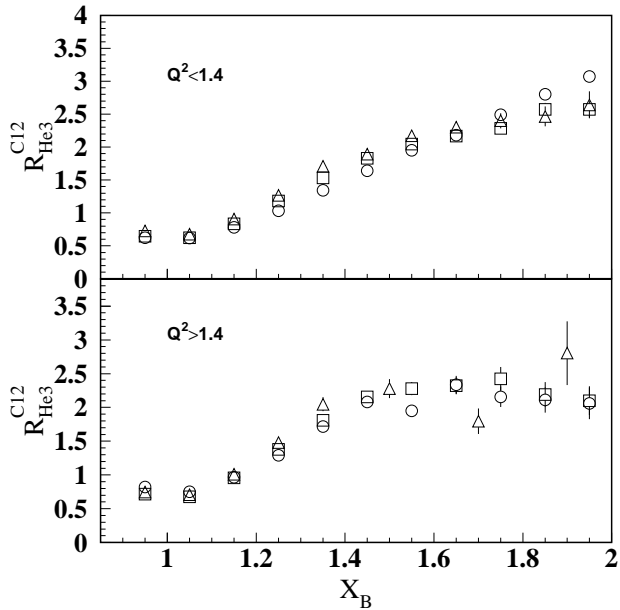}
\caption[]{$R^C_{^3He}$, the per-nucleon yield ratios for $^{12}$C to $^3$He. 
a) $\circ$ $0.65<$Q$^2<0.85$; $\square$ $0.9<$Q$^2<1.1$; 
$\triangle$  $1.15<$Q$^2<1.35$ (GeV/c)$^2$, all at incident energy 2.261 GeV.

b) $\circ$ $1.4<$Q$^2<1.6$ (GeV/c)$^2$ and at incident energy 2.261 GeV; 
$\square$ $1.4<$Q$^2<2.0$ ; $\triangle$ $2.0<$Q$^2<2.6$  (GeV/c)$^2$, 
both at incident energy 4.461 GeV. Statistical errors are shown only.}
\label{fig:Cratio}
\end{center}
\end{figure}
\begin{figure}[ht]
\begin{center}
\epsfig{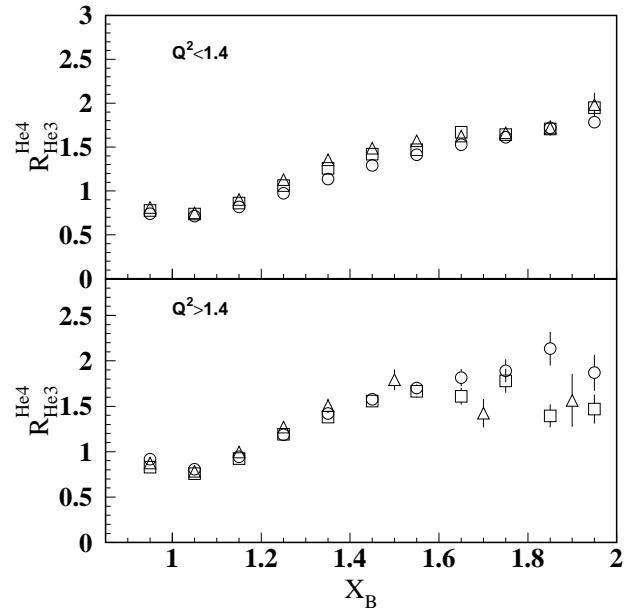}
\caption[]{The same as   Fig.~\ref{fig:Cratio} for $^{4}$He.}  
\label{fig:Heratio}
\end{center}
\end{figure} 
\begin{figure}[ht]
\begin{center}
\epsfig{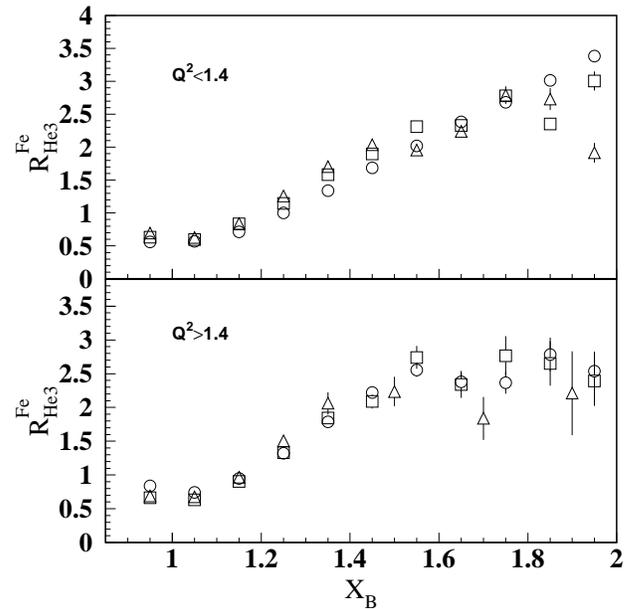}
\caption[]{The same as   Fig.~\ref{fig:Cratio} for $^{56}$Fe. }  
\label{fig:Feratio}
\end{center}
\end{figure}

\subsection {\large \bf  Radiative Corrections }
The cross section ratios were corrected for radiative
effects. The radiative correction for each target as a function of
Q$^2$ and $x_B$ was calculated as the
ratio
\begin{eqnarray}
C_{rad}(x_B,Q^2)= \frac{d\sigma^{rad}(x_B,Q^2)}{d\sigma^{norad}(x_B,Q^2)},
\label{RadCor}
\end{eqnarray}
where $d\sigma^{rad}(x_B,Q^2)$ and $d\sigma^{norad}(x_B,Q^2)$ are the
radiatively corrected and uncorrected theoretical cross sections,
respectively. The cross sections have been calculated using
\cite{Misak2}. 

\section {\large \bf RESULTS}
We constructed ratios of normalized, acceptance- and radiative-corrected
inclusive electron yields on nuclei $^{4}$He, $^{12}$C and $^{56}$Fe 
divided the yield on $^3$He in the range of
kinematics 1$<x_B <$2. Assuming that electron detection efficiency from
different targets is the same, these ratios, weighted by atomic number, are
equivalent to the ratios of cross sections in Eq.~\ref{ratio1a}. 

The normalized yields for each $x_B$ and Q$^2$ 
bin have been calculated as:
\begin{eqnarray}
\frac {d \cal Y}{dQ^2dx_B}~=~\frac {N_{e'}}{\Delta Q^2\Delta x_B
N_eN_T}\cdot \frac{1}{Acc} 
\label{CrSec}
\end{eqnarray}
where $N_e$ and $N_T$ are the number of electrons and target nuclei, respectively, 
Acc is the acceptance correction factor, and 
$\Delta$Q$^2$ and $\Delta x_B$ are the bin sizes in Q$^2$ and in $x_B$,
respectively. Since electron detection efficiency in CLAS is expected to be $>$
96\%, we compare obtained yields with radiated cross sections calculated by  
Ref.~\cite{Misak2} code. Within systematic uncertainies (see below) the 
satisfactory agreement has been found 
between our results and the calculations from Ref.~\cite{Misak2} that were tuned on SLAC
~\cite{Day2} and Jefferson Lab Hall C ~\cite{Petit} data.

The ratios $R^A_{^3He}$, also corrected for radiative effects, are
defined as:
\begin{eqnarray}
R^A_{^3He}(x_B)~=~\frac {3{\cal Y^A}}{A{\cal Y}^{^3 He}} \frac {C_{Rad}^A}{C_{Rad}^{^3He}}, 
\label{RatY}
\end{eqnarray}
where $\cal Y$ is the normalized yield in a given ($x_B$, Q$^2$) bin,  and $C_{Rad}$
is the radiative correction factor from Eq. \ref{RadCor} for each nucleus. 

Fig. \ref{fig:Cratio} shows these ratios for $^{12}$C 
at several values of Q$^2$.  Figs. \ref{fig:Heratio} and
\ref{fig:Feratio} 
show these ratios for $^4$He and  $^{56}$Fe, respectively.
These data have the following important characteristics:
\begin{itemize}
\item There is a clear Q$^2$ evolution of the shape of ratios:
\begin{itemize}
\item At low Q$^2$ (Q$^2<$1.4 (GeV/c)$^2$), R$^A_{^3He}(x_B)$ increases
with ${x_B}$ in the entire 1$<x_B<$2 range (see Figs. \ref{fig:Cratio}a -- 
\ref{fig:Feratio}a).
\item At high Q$^2$ (Q$^2 \geq $1.4 (GeV/c)$^2$) R$^A_{^3He}(x_B)$ is
independent of $x_B$ for $x_B > x_B^0 \approx$1.5. (See
Figs. \ref{fig:Cratio}b -- \ref{fig:Feratio}b.)
\end{itemize}
\item The value of R$^A_{^3He}(x_B)$  in the scaling regime is
independent of Q$^2$. 
\item The value of R$^A_{^3He}(x_B)$ in the scaling regime for $A >$ 10 
suggests a weak dependence on target mass.
\end{itemize}

\subsection {\large \bf Systematic Errors } 
The systematic errors in this measurement are different for
different targets and include uncertainties in:
\begin{itemize}
\item fiducial cut applied:  $\approx$ 1\%; 
\item  radiative correction factors:  $\approx$ 2\%;
\item target densities and thicknesses: $\approx$ 0.5\% and 1.0\% for
solid targets; 0.5\% and 3.5\% for liquid targets, respectively.
\item acceptance correction factors (Q$^2$ dependent): between 2.2 and 4.0\%
for solid targets and between 1.8 and 4.3\%  for liquid targets.
\end{itemize}
Some of systematic  uncertainties will cancel out in the yield ratios.  
For the $^{3}$He/$^{4}$He ratio, all uncertainties except those on the beam 
current and
the target density divide out, giving a total systematic uncertainty
of 0.7\%.  
For the solid-target to $^3$He ratios, only the electron detection
efficiency cancels. The quadratic sum of the other uncertainties is
between 5\% and 7\%, depending on Q$^2$. The systematic uncertainties on the
ratios for all targets and Q$^2$ are presented in Table~\ref{systerror}.
\begin {table} [ht]
\begin{center}
\begin{tabular} {|c||c||c||c||c|} \hline

  Q$^2$     &~~~~1.55~~~~&~~~~1.85~~~~&~~~~2.15~~~~&~~~~2.45~~~~  \\ \hline 

  $\delta R^{A}$ &  7.1 &  5.8 & 4.9&  5.1   \\ \hline 
  $\delta R^{He4}$ &  0.7 &  0.7  & 0.7 &  0.7    \\ \hline 

\end{tabular}
\caption {Systematic uncertainties $\delta R^A$ and $\delta R^{He4}$ for the
 ratio of normalized inclusive yields $R^{C,Fe}_{^3He}$ and $R^{^4He}_{^3He}$
 respectively, in \%. Q$^2$ in (GeV/c)$^2$, and 
$\Delta$Q$^2=\pm$0.15 (GeV/c)$^2$.}
\label{systerror}
\end{center}
\end{table}

\subsection {\large \bf Probabilities of Two-Nucleon Short Range
 Correlations in Nuclei}  
Our data are clearly consistent
with the predictions of the $NN$ SRC model. 
The  obtained ratios, R$^A_{^3He}$, for $ 1.4< $Q$^2 <2.6$  (GeV/c)$^2$ region 
are shown in Table II as a function of $x_B$. Fig.~\ref{fig:ratio1bin} shows 
these ratios for the $^{12}$C and $^{56}$Fe targets  together with the SRC 
calculation results using of Ref.~ \cite{Misak2} which used 
the estimated scaling factors, $a_2(A)$, (per-nucleon probabilitiy of $NN$ SRC 
in nucleus A) from Ref.~\cite{FSDS}. Good agreement between our data and 
calculation is seen. Note that one of the goals of the present paper is to determine these 
factors more precisely (see below). 
\begin{figure}[ht]
\begin{center}
\epsfig{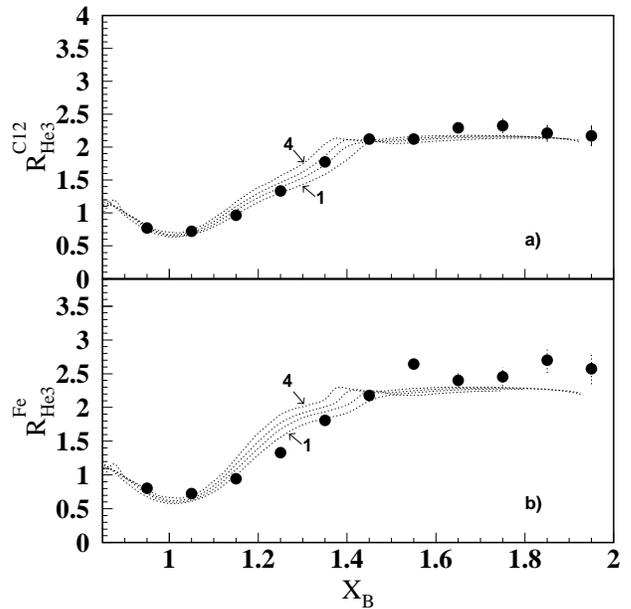}
\caption[]{$R^A_{^3He}(x_B)$ as a function of $x_B$ for 1.4$<$Q$^2<$2.6 
(GeV/c)$^2$, statistical errors are shown only. Curves are SRC model predictions for 
different Q$^2$ in the range  1.4 (GeV/c)$^2$ (curve 1) to 2.6 (GeV/c)$^2$ (curve 4), 
respectively, for 
 a) $^{12}$C, b) $^{56}$Fe.}
\label{fig:ratio1bin}
\end{center}
\end{figure} 
\begin{figure}[ht]
\begin{center}
\epsfig{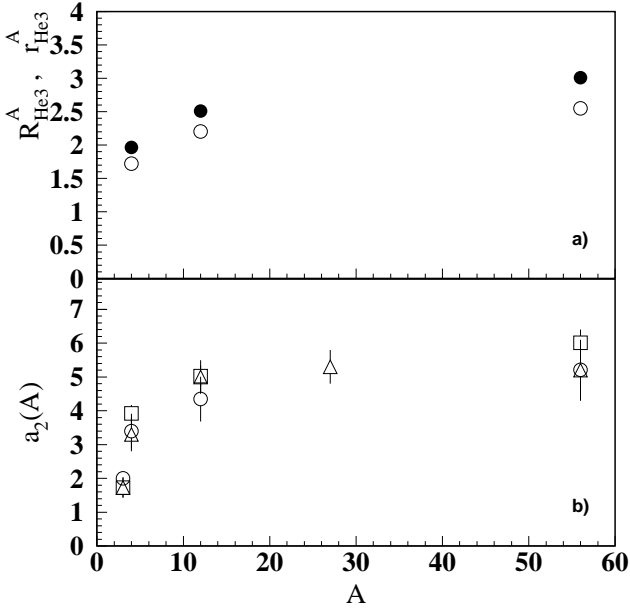}
\caption[]{a) R$^A_{^3He}$ ($\circ$) and $r^A_{^3He}$ ($\bullet$) versus $A$.
b) a$_2$(A) versus $A$. $\circ$ - a$_2$(A) is obtaned from  Eq.(\ref{a2_3})
using the experimental value of a$_2$(3) Ref.~\cite{FSDS}; $\square$ - a$_2$(A) 
is obtaned using the theoretical value of a$_2$(3). For errors shown see  
caption of Table III. $\triangle$ - data from Ref.~\cite{FSDS}.}
\label{fig:a2a}
\end{center}
\end{figure} 

Experimental data in the scaling region can be used to estimate the relative 
probabilities of $NN$ SRC in nuclei compared to $^3$He.
According to Ref.~\cite{FSreps} the ratio of these probabilities is proportional
to:
\begin{eqnarray}
r^A_{^3He} = \frac {(2\sigma_p+\sigma_n)\sigma_A}{(Z\sigma_p+N\sigma_n)\sigma_{3He}},
\label{a2_1}
\end{eqnarray}
where $\sigma_A$ and $\sigma_{3He}$ are the $A(e,e')$ and
$^3$He$(e,e')$ inclusive cross sections  respectively.  $\sigma_p$
and $\sigma_n$ are the electron-proton and electron-neutron elastic 
scattering cross
sections respectively.  $Z$ and $N$ are the number of protons and
neutrons in nucleus $A$. 
Using Eq.~(\ref{ratio1a}) the ratio of Eq.~(\ref{a2_1}) can be 
related to the experimentally measured ratios R$^A_{^3He}$ as
\begin{eqnarray}
r^A_{^3He} = R^A_{^3He}(x_B,Q^2)\times \frac 
{A(2\sigma_p+\sigma_n)}
{3(Z\sigma_p+N\sigma_n)}
\label{a2_2}
\end{eqnarray}

\begin {table} [ht]
\begin{center}
\small{
\begin{tabular} {|c||c||c||c|} \hline
  $X_B$             & $^4$He          & $^{12}$C       & $^{56}$Fe      \\ \hline 
  0.95 $\pm$ 0.05   & 0.86  $\pm$ 0.004 & 0.77 $\pm$ 0.003 & 0.80 $\pm$ 0.004  \\ \hline 
  1.05 $\pm$ 0.05   & 0.78  $\pm$ 0.004 & 0.72 $\pm$ 0.003 & 0.72 $\pm$ 0.004  \\ \hline 
  1.15 $\pm$ 0.05   & 0.94  $\pm$ 0.006 & 0.96 $\pm$ 0.006 & 0.94 $\pm$ 0.007  \\ \hline 
  1.25 $\pm$ 0.05   & 1.19  $\pm$ 0.012 & 1.33 $\pm$ 0.012 & 1.33 $\pm$ 0.015  \\ \hline 
  1.35 $\pm$ 0.05   & 1.41  $\pm$ 0.021 & 1.77 $\pm$ 0.025 & 1.81 $\pm$ 0.030  \\ \hline 
  1.45 $\pm$ 0.05   & 1.58  $\pm$ 0.033 & 2.12 $\pm$ 0.044 & 2.17 $\pm$ 0.055  \\ \hline 
  1.55 $\pm$ 0.05   & 1.71  $\pm$ 0.049 & 2.12 $\pm$ 0.059 & 2.64 $\pm$ 0.087  \\ \hline 
  1.65 $\pm$ 0.05   & 1.70  $\pm$ 0.063 & 2.29 $\pm$ 0.085 & 2.40 $\pm$ 0.109  \\ \hline 
  1.75 $\pm$ 0.05   & 1.85  $\pm$ 0.089 & 2.32 $\pm$ 0.110 & 2.45 $\pm$ 0.139  \\ \hline 
  1.85 $\pm$ 0.05   & 1.65  $\pm$ 0.100 & 2.21 $\pm$ 0.128 & 2.70 $\pm$ 0.190  \\ \hline 
  1.95 $\pm$ 0.05   & 1.71  $\pm$ 0.124 & 2.17 $\pm$ 0.157 & 2.57 $\pm$ 0.227  \\ \hline 

\end{tabular}}
\caption {The ratios, $R^A_{^3He}$,  measured in 1.4$<$Q$^2<$2.6 (GeV/c)$^2$ 
interval. Errors are statistical only.}
\label{ratios}
\end{center}
\end{table}

To obtain the numerical values for r$^A_{^3He}$ we calculated the second 
factor in Eq.~(\ref{a2_2}) using
parameterizations for the neutron and proton form factors \cite{Ina,
KDJ, Bosted, Miller}. We found the average values of these factors to be
1.14$\pm$0.02 for $^4$He and $^{12}$C, and 1.18$^\pm$0.02 for
$^{56}$Fe. Note that these
factors vary slowly over our Q$^2$ range. 
For r$^A_{^3He}$  calculations the 
experimental data were integrated
over Q$^2>$1.4~(GeV/c)$^2$ and $x_B>$1.5 for each nucleus.  
The ratio of integrated
yields, R$^A_{^3He}(x_B)$, are presented in the first column of
Table~\ref{ratios} and in Fig.~\ref{fig:a2a}a (rectangles).
The ratios r$^A_{^3He}$ are shown in the second column of Table~\ref{ratios}
and by the circles in Fig.~\ref{fig:a2a}a.  
One can see that the ratios $r^A_{^3He}$
are 2.5 - 3.0 for $^{12}$C and $^{56}$Fe, and approximately 1.95 for $^4$He.
\begin{figure}[ht]
\begin{center}
\epsfig{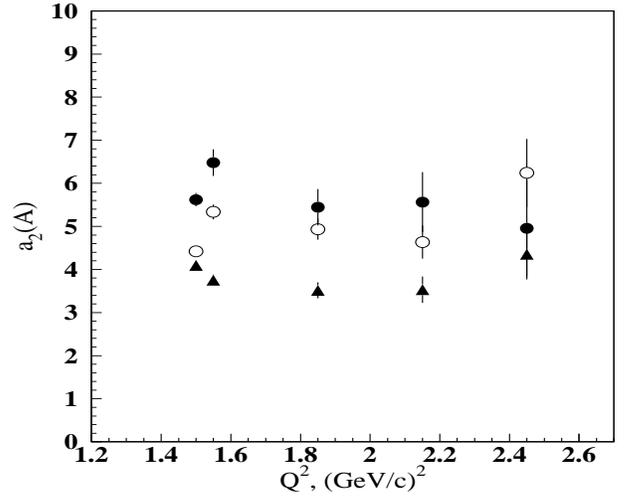}
\caption[]{Q$^2$ - dependences of $a_2(A)$ parameters obtained by
multiplying r$^A_{^3He}$ (=$\frac {a_2(A)}{a_2(3)}$) with the  
theoretical values of $a_2(3)$. For errors shown see caption of Table III.
$\blacktriangle$ $^4$He, $\circ$ $^{12}$C and $\bullet$ $^{56}$Fe.}
\label{fig:Fig.7a}
\end{center}
\end{figure} 

The per-nucleon SRC probability in nucleus $A$ relative to $^3$He is 
proportional to r$^A_{^3He}\sim$ a$_2$(A)/a$_2$(3), where a$_2$(A) and a$_2$(3)
are the per-nucleon probability of SRC relative to deuterium for nucleus $A$ and
$^3$He. As was discussed earlier, the direct relation of r$^A_{^3He}$ to
the per-nucleon probabilities of SRC has an uncertainty of up to 20\% due to pair
center-of-mass motion. Within this uncertainty, we will define the per-nucleon SRC 
probabilities of nuclei relative
to deuterium as:
\begin{eqnarray}
a_2(A)~=~r^A_{^3He} \cdot a_2(3)
\label{a2_3}
\end{eqnarray}

Two values of a$_2$(3) have been used to calculate a$_2$(A). First is the
experimentally obtained value from Ref.~\cite{FSDS}, a$_2$(3)=1.7$\pm$ 0.3, and the
second the value from the calculation using the wave function for deuterium and
$^3$He, a$_2$(3)=2$\pm $0.1. 
Similar results were obtained in Ref.~\cite{jdeforest}. 
\begin {table} [ht]
\begin{center}
\small{
\begin{tabular} {|c||c||c||c||c||c|} \hline
      & R$^A_{^3He}(x_B)$ & r$^A_{^3He}$ & a$_2$(A)$^{exp}$ & a$_2$(A$)^{theor}$  \\ \hline 
$^{4}$He & 1.72 $\pm$ 0.03 & 1.96 $\pm$ 0.05 & 3.39 $\pm$ 0.51 & 3.93 $\pm$ 0.24 \\ \hline 
$^{12}$C & 2.20 $\pm$ 0.04 & 2.51 $\pm$ 0.06 & 4.34 $\pm$ 0.66 & 5.02 $\pm$ 0.31 \\ \hline 
$^{56}$Fe & 2.54 $\pm$ 0.06 & 3.00 $\pm$ 0.08 & 5.21 $\pm$ 0.79 & 6.01 $\pm$ 0.38  \\ \hline 
\end{tabular}}
\caption {R$^A_{^3He}(x_B)$ is the ratio of normalized $(e,e')$ yields 
for nucleus $A$ to $^3$He. $r^A_{^3He}$ is the relative per-nucleon
probability of SRC for the two nuclei.  Both values are obtained from
the scaling region (Q$^2>$1.4 (GeV/c)$^2$ and x$_B>$1.5). a$_2$(A)$^{exp}$ 
and a$_2$(A)$^{theor}$ are the a$_2$(A) parameters obtained by
multiplying r$^A_{^3He}$ (=$\frac {a_2(A)}{a_2(3)}$) with the experimental and/or 
theoretical values of $a_2(3)$. The R$^A_{^3He}(x_B)$ ratios 
include statistical errors only.  
Errors of r$^A_{^3He}$ include also uncertainties in the second factor 
in Eq.(\ref{a2_2}). 
Errors in a$_2$(A)$^{exp}$ and a$_2$(A)$^{theor}$ 
also include uncertainties in the corresponding values of a$_2$(3). 
For systematic 
uncertainties see Table~\ref{systerror}. There is an overall theoretical
uncertainty of 20\% in converting these ratios into SRC probabilities.}
\label{ratios}
\end{center}
\end{table}

The per-nucleon probability of SRC for nucleus $A$ relative to
deuterium is shown in the third and fourth columns of
Table~\ref{ratios} and in Fig.~\ref{fig:a2a}b. The uncertainties in
existing a$_2$(3) are included in the errors for a$_2$(A). The
results from Ref.~\cite{FSDS} are shown as well.  One can see that
a$_2$(A) changes significantly from $A=4$ to $A = 12$ but does not
change significantly for $A \ge 12$.  There are approximately 5.5 times
as much SRC for $A\geq $12 than for deuterium, and approximately 3.5 times
as much SRC for $^4$He as for deuterium.  These results are consistent
with the analysis of the previous SLAC $(e,e')$ data \cite{FSDS}. 
They are also consistent with calculations of Ref.~\cite{jdeforest}.

Fig.~\ref{fig:Fig.7a} shows the measured Q$^2$ dependence of the relative SRC
probability, a$_2$(A), which appears to be Q$^2$ independent for all targets.    
\section {\large \bf SUMMARY}
The $A(e,e')$ inclusive electron scattering cross section ratios 
of $^4$He, $^{12}$C, and $^{56}$Fe to $^3$He have been
measured for the first time under identical kinematical conditions.

It is shown that: 
\begin{enumerate}
\item These ratios are independent of $x_B$  (scale) for $x_B >$ 1.5 and
Q$^2>$1.4 (GeV/c)$^2$, {\it i.e.} for high recoil momentum.  
The ratios do not scale for Q$^2<$1.4 (GeV/c)$^2$.
\item These ratios in the scaling region are independent of Q$^2$, and 
approximately independent of $A$ for $A\ge 12$
\item These features were predicted by the Short Range Correlation model of
inclusive $A(e,e^{\prime})$ scattering at large $x_B$, and consistent with the 
kinematical expectation that two nucleon short range correlations are dominating 
in the nuclear wave function at $p_m\gtrsim$ 300 MeV/c \cite{FSDS}.
\item The observed scaling shows that momentum distributions at high-momenta have 
the same shape for all nuclei and differ only by a scale factor.
\item Using the SRC model, the values of the ratios in the scaling region
were used to derive the relative probabilities of SRC in nuclei
compared to deuterium.  The per-nucleon probability of Short Range
Correlations in nuclei relative to deuterium is $\approx$ 3.5 times larger for
$^4$He and 5.0 - 5.5 times larger for $^{12}$C and $^{56}$Fe.
\end{enumerate}
\vskip 1.cm
\centerline {\large \bf ACKNOWLEDGMENT}
\vskip 0.5cm 
We acknowledge the efforts of the staff of the Accelerator and Physics
Divisions (especially the Hall B target group) at Jefferson Lab in
their support of this experiment.  We also
acknowledge useful discussions with D. Day and E. Piasetzki.  
This work was supported in part by the U.S. Department of Energy, 
the National Science Foundation, the French
Commissariat a l'Energie Atomique, the French Centre National de la 
Recherche Scientifique, the Italian Istituto Nazionale di
Fisica Nucleare, and the Korea Research Foundation. U. Thoma acknowledges 
an ``Emmy Noether'' grand from the Deutsche Forschungsgemeinschaft. 
The Souteastern Universities Research Association (SURA) operates the 
Thomas Jefferson National Acceleraror Facility for the United States 
Department of Energy under contract DE-AC05-84ER40150.  

\normalsize
\end{document}

%% file: authorE2.tex
\newcommand*{\YEREVAN }{ Yerevan Physics Institute, Yerevan 375036 , Armenia} \affiliation{\YEREVAN } 
\newcommand*{\ASU }{ Arizona State University, Tempe, Arizona 85287-1504} \affiliation{\ASU } 
\newcommand*{\UCLA }{ University of California at Los Angeles, Los Angeles, California  90095-1547} 
\affiliation{\UCLA } 
\newcommand*{\CMU }{ Carnegie Mellon University, Pittsburgh, Pennsylvania 15213} \affiliation{\CMU } 
\newcommand*{\CUA }{ Catholic University of America, Washington, D.C. 20064} \affiliation{\CUA } 
\newcommand*{\SACLAY }{ CEA-Saclay, Service de Physique Nucl\'eaire, F91191 Gif-sur-Yvette, Cedex, France} 
\affiliation{\SACLAY } 
\newcommand*{\CNU }{ Christopher Newport University, Newport News, Virginia 23606} \affiliation{\CNU } 
\newcommand*{\UCONN }{ University of Connecticut, Storrs, Connecticut 06269} \affiliation{\UCONN } 
\newcommand*{\DUKE }{ Duke University, Durham, North Carolina 27708-0305} \affiliation{\DUKE } 
\newcommand*{\GBEDINBURGH }{ Edinburgh University, Edinburgh EH9 3JZ, United Kingdom} \affiliation{\GBEDINBURGH } 
\newcommand*{\FIU }{ Florida International University, Miami, Florida 33199} \affiliation{\FIU } 
\newcommand*{\FSU }{ Florida State University, Tallahassee, Florida 32306} \affiliation{\FSU } 
\newcommand*{\GWU }{ The George Washington University, Washington, DC 20052} \affiliation{\GWU } 
\newcommand*{\GBGLASGOW }{ University of Glasgow, Glasgow G12 8QQ, United Kingdom} \affiliation{\GBGLASGOW } 
\newcommand*{\INFNFR }{ INFN, Laboratori Nazionali di Frascati, Frascati, Italy} \affiliation{\INFNFR } 
\newcommand*{\INFNGE }{ INFN, Sezione di Genova, 16146 Genova, Italy} \affiliation{\INFNGE } 
\newcommand*{\ORSAY }{ Institut de Physique Nucleaire ORSAY, Orsay, France} \affiliation{\ORSAY } 
\newcommand*{\BONN }{ Institute f\"{u}r Strahlen und Kernphysik, Universit\"{a}t Bonn, Germany} \affiliation{\BONN } 
\newcommand*{\ITEP }{ Institute of Theoretical and Experimental Physics, Moscow, 117259, Russia} \affiliation{\ITEP } 
\newcommand*{\JMU }{ James Madison University, Harrisonburg, Virginia 22807} \affiliation{\JMU } 
\newcommand*{\KYUNGPOOK }{ Kungpook National University, Taegu 702-701, South Korea} \affiliation{\KYUNGPOOK } 
\newcommand*{\MIT }{ Massachusetts Institute of Technology, Cambridge, Massachusetts  02139-4307} \affiliation{\MIT } 
\newcommand*{\UMASS }{ University of Massachusetts, Amherst, Massachusetts  01003} \affiliation{\UMASS } 
\newcommand*{\UNH }{ University of New Hampshire, Durham, New Hampshire 03824-3568} \affiliation{\UNH } 
\newcommand*{\NSU }{ Norfolk State University, Norfolk, Virginia 23504} \affiliation{\NSU } 
\newcommand*{\OHIOU }{ Ohio University, Athens, Ohio  45701} \affiliation{\OHIOU } 
\newcommand*{\ODU }{ Old Dominion University, Norfolk, Virginia 23529} \affiliation{\ODU } 
\newcommand*{\PENST}{Pennsylvania State University, State Collage, Pennsylvania 16802} \affiliation{\PENST }
\newcommand*{\PITT }{ University of Pittsburgh, Pittsburgh, Pennsylvania 15260} \affiliation{\PITT } 
\newcommand*{\ROMA }{ Universita' di ROMA III, 00146 Roma, Italy} \affiliation{\ROMA } 
\newcommand*{\RPI }{ Rensselaer Polytechnic Institute, Troy, New York 12180-3590} \affiliation{\RPI } 
\newcommand*{\RICE }{ Rice University, Houston, Texas 77005-1892} \affiliation{\RICE } 
\newcommand*{\URICH }{ University of Richmond, Richmond, Virginia 23173} \affiliation{\URICH } 
\newcommand*{\SCAROLINA }{ University of South Carolina, Columbia, South Carolina 29208} \affiliation{\SCAROLINA } 
\newcommand*{\UTEP }{ University of Texas at El Paso, El Paso, Texas 79968} \affiliation{\UTEP } 
\newcommand*{\JLAB }{ Thomas Jefferson National Accelerator Facility, Newport News, Virginia 23606} 
\affiliation{\JLAB } 
\newcommand*{\UNIONC }{Union College, Schenectady, NY 12308} \affiliation{\UNIONC } 
\newcommand*{\VT }{ Virginia Polytechnic Institute and State University, Blacksburg, Virginia   24061-0435} 
\affiliation{\VT } 
\newcommand*{\VIRGINIA }{ University of Virginia, Charlottesville, Virginia 22901} \affiliation{\VIRGINIA } 
\newcommand*{\WM }{ College of William and Mary, Williamsburg, Virginia 23187-8795} \affiliation{\WM } 

\newcommand*{\NOWNCATU }{ North Carolina Agricultural and Technical State University, Greensboro, NC 27411}
\newcommand*{\NOWGBGLASGOW }{ University of Glasgow, Glasgow G12 8QQ, United Kingdom}
\newcommand*{\NOWJLAB }{ Thomas Jefferson National Accelerator Facility, Newport News, Virginia 23606}
\newcommand*{\NOWSCAROLINA }{ University of South Carolina, Columbia, South Carolina 29208}
\newcommand*{\NOWFIU }{ Florida International University, Miami, Florida 33199}
\newcommand*{\NOWINFNFR }{ INFN, Laboratori Nazionali di Frascati, Frascati, Italy}
\newcommand*{\NOWOHIOU }{ Ohio University, Athens, Ohio  45701}
\newcommand*{\NOWCMU }{ Carnegie Mellon University, Pittsburgh, Pennsylvania 15213}
\newcommand*{\NOWINDSTRA }{ Systems Planning and Analysis, Alexandria, Virginia 22311}
\newcommand*{\NOWASU }{ Arizona State University, Tempe, Arizona 85287-1504}
\newcommand*{\NOWCISCO }{ Cisco, Washington, DC 20052}
\newcommand*{\NOWUK }{ Kentucky, LEXINGTON, KENTUCKY 40506}
\newcommand*{\NOWSACLAY }{ CEA-Saclay, Service de Physique Nucl\'eaire, F91191 Gif-sur-Yvette, Cedex, France}
\newcommand*{\NOWRPI }{ Rensselaer Polytechnic Institute, Troy, New York 12180-3590}
\newcommand*{\NOWDUKE }{ Duke University, Durham, North Carolina 27708-0305}
\newcommand*{\NOWUNCW }{ North Carolina}
\newcommand*{\NOWHAMPTON }{ Hampton University, Hampton, VA 23668}
\newcommand*{\NOWTulane }{ Tulane University, New Orleans, Lousiana  70118}
\newcommand*{\NOWKYUNGPOOK }{ Kungpook National University, Taegu 702-701, South Korea}
\newcommand*{\NOWCUA }{ Catholic University of America, Washington, D.C. 20064}
\newcommand*{\NOWGEORGETOWN }{ Georgetown University, Washington, DC 20057}
\newcommand*{\NOWJMU }{ James Madison University, Harrisonburg, Virginia 22807}
\newcommand*{\NOWURICH }{ University of Richmond, Richmond, Virginia 23173}
\newcommand*{\NOWCALTECH }{ California Institute of Technology, Pasadena, California 91125}
\newcommand*{\NOWMOSCOW }{ Moscow State University, General Nuclear Physics Institute, 119899 Moscow, Russia}
\newcommand*{\NOWVIRGINIA }{ University of Virginia, Charlottesville, Virginia 22901}
\newcommand*{\NOWRICE }{ Rice University, Houston, Texas 77005-1892}
\newcommand*{\NOWINFNGE }{ INFN, Sezione di Genova, 16146 Genova, Italy}
\newcommand*{\NOWBATES }{ MIT-Bates Linear Accelerator Center, Middleton, MA 01949}
\newcommand*{\NOWODU }{ Old Dominion University, Norfolk, Virginia 23529}
\newcommand*{\NOWVSU }{ Virginia State University, Petersburg,Virginia 23806}
\newcommand*{\NOWUNIONC }{Union College, Schenectady, NY 12308} 
\newcommand*{\NOWORST }{ Oregon State University, Corvallis, Oregon 97331-6507}
\newcommand*{\NOWCNU }{ Christopher Newport University, Newport News, Virginia 23606}
\newcommand*{\NOWGWU }{ The George Washington University, Washington, DC 20052}
\author{K.Sh.~Egiyan}     \affiliation{\YEREVAN}
\author{N.Dashyan}     \affiliation{\YEREVAN}
\author{M.~Sargsian}   \affiliation{\FIU} 
\author{S.~Stepanyan}   \affiliation{\JLAB} \altaffiliation[Current address:]{\NOWODU}
\author{L.B.~Weinstein}     \affiliation{\ODU}
\author{G.~Adams}\affiliation{\RPI}
\author{P.~Ambrozewicz}\affiliation{\FIU}
\author{E.~Anciant}     \affiliation{\SACLAY}
\author{M.~Anghinolfi}     \affiliation{\INFNGE}
\author{B.~Asavapibhop}     \affiliation{\UMASS}
% ...
\collaboration{The CLAS Collaboration}     \noaffiliation